\documentclass[12pt.a4]{article}

\usepackage{latexsym}
\usepackage{amsmath}
\usepackage{bm}
\usepackage{txfonts}
\title{Singularities of Magnetic Monopoles\\
 for Dirac and 't Hooft-Polyakov Theories\\ 
by Pre-potential Method}
\author{Masakatsu Kenmoku\thanks{m.kenmoku@cc.nara-wu.ac.jp}\\
Nara Women's University, Nara 630-8506, Japan}
\date{\empty}

\begin{document}
\maketitle
\abstract{

 The magnetic monopole is one of the important problems 
in the early stage of universe as well as observations
and experiments on Earth. 
We study the existence or non-existence of the 
Dirac and the 't Hooft-Polyakov magnetic monopole theories 
using the pre-potential $\boldsymbol{C}$, which is defined to derive  
the vector potential by the curl operation as 
$\boldsymbol{A}=\nabla \times \boldsymbol{C}$ . 
We assert that the magnetic singularity exists for the 't Hooft-Polyakov monopole   
in SO(3) gauge theory, as well as for the Dirac monopole in U(1) gauge theory.  
The regularization method confirms our assertion. 
}


\section{Introduction}
\label{intro}

Existence or non-existence  of magnetic monopoles is
 one of the most interesting problems in both theoretical and observational physics.   

In standard Maxwell electrodynamics, magnetic monopoles do not exist analytically.  
P. A. M. Dirac found the magnetic monopole solution in Maxwell's $U(1)$ gauge theory 
in order to make the theory symmetric between electricity and magnetism by introducing a semi-infinite string singularity, whose edge plays the role of the magnetic monopole \cite{Dirac1931}. 
  
Motivated by the Dirac's monopole theory and 
the quantized magnetic flux  lines in a superconductor 
by the work of Nielsen and Olsen \cite{NielsenOlsen}, 
't Hooft and Polyakov derived the topological magnetic monopole without singularities in $SO(3)$ gauge theory 
\cite{'t Hooft1974, Polyakov1974}.  
We study the existence or non-existence of the singularity in the Dirac and 
the 't Hooft-Polyakov monopole theories using our pre-potential method. 
Singularities are difficult to find, because they occur at the boundary of analytic function and are expressed as a general function or distribution ( $\delta$ function, $\theta$ function,...).  
The pre-potential method is useful for finding singularities in monopole theories. 

The paper is organized as follows: 
In section.2, the pre-potential method for the existence of monopoles and singularities is studied.   
In subsection 2.1, 
the explicit form of the Dirac string singularity is derived in a manifest way 
using our pre-potential method. 
 In subsection 2.2, the 't Hooft-Polyakov monopole theories are studied using our pre-potential method 
and the singularity in the local Cartesian coordinate is derived.  
In section 3, the regularization method is developed  
 to confirm the existence of the monopole singularities by our pre-potential method. 
In section 4, the gauge transformation between the Dirac and the 't Hooft-Polyakov monopoles 
is studied. Conclusion and remarks are given in section 5.  

One of the motivations to the monopole research is the electromagnetic monopole in the standard model by Cho and Maison \cite{Cho1997}.
Preliminary work of this topic was presented at the international conference ICGAC-15 \cite{Kenmoku2023}. 

In the following, we use the metric ${\rm diag}(\eta)=(-1,1,1,1)$ and the natural unit $\hbar =c=1$.


\section{The pre-potential method and the singularity for monopoles }
\renewcommand{\theequation}{\thesection,\arabic{equation}}
\setcounter{equation}{0}

In this section, we will study the existence or non-existence of monopoles and singularities 
for the Dirac (D) and the 't Hooft-Polyakov (HP) theories using the pre-potential method. 

\subsection{The pre-potential method and the singularity for the Dirac monopole }

P. A. M. Dirac studied $U(1)$ gauge theory and obtained the magnetic monopole solution for the gauge potential  
\[
{\boldsymbol{A}^D}=\frac{g (-y, x, 0)}{r(r+z)},
\]
where $g$ denotes the magnetic charge. 
The Dirac string is along the negative $z$ axis $(r+z=0)$ 
and the magnetic monopole appears at its end. 
We will study the Dirac string singularity  using the pre-potential method. 

The pre-potential  
for the Dirac monopole is introduced as,  
\begin{eqnarray}
{\boldsymbol{C}}^{D} := -g(0, 0, {\ln(r+z)}). 
\end{eqnarray} 
The gauge potential of the Dirac monopole solution is obtained from the curl operation to $\boldsymbol {C}^{D}$ as, 
\begin{equation}
{\boldsymbol{A}}^{D} := \nabla {\times} {\boldsymbol{C}^{D}} =\frac{g (-y, x, 0)}{r(r+z)}.
\end{equation}
The magnetic field is given by the gauge potential:  
\begin{equation}
{\boldsymbol{B}}^{D}=\nabla \times {\boldsymbol{A}}^D =\nabla \times (\nabla \times {\boldsymbol{C}}^D)
=\nabla (\nabla \cdot {\boldsymbol{C}}^D)-\Delta {\boldmath{C}}^D.  
\end{equation} 
The first term of the magnetic field is of the Coulomb pole type  
\begin{equation}
{\boldsymbol{B}}^D ( {\rm pole}):= \nabla (\nabla \cdot {\boldsymbol C}^D) = \frac{g}{r^2}  {{\rm{e}}}_r , 
\, \,  { {\rm{e}}}_r =(x, y, z)/r ,
\end{equation}
and the magnetic charge of the pole type is 
\begin{equation}
\rho^D({\rm pole}):=
\nabla
\cdot{\boldsymbol{B}}^D ({\rm pole}) = g\nabla \frac{g}{r^2}  {{\rm{e}}}_r = 4\pi g \delta^{(3)}({\boldsymbol r}) . 
\end{equation}
The second singular term of the magnetic field 
\[ {\boldsymbol B}^D (\rm singular):= -\Delta {\boldsymbol C}^D \]
gives the magnetic charge:  
\begin{equation}
\rho^D( {\rm singular}):= \nabla \cdot {\boldsymbol B}^D ({\rm singular}) 
=g  \partial_z \Delta {\rm{ln}}(r+z) =g \Delta \frac{1}{r} =-4\pi  g \delta^{(3)}({\boldsymbol{r}}).  
\end{equation}
The sum of $\rho^D( {\rm pole})$ and $\rho^D( {\rm singular})$ gives the zero total magnetic charge. 

The string singularity is therefore included in the magnetic field as 
\begin{equation}
{\boldsymbol B}^D=\frac{g}{r^2}{\boldsymbol e}_r+4\pi g \delta (x) \delta (y) \theta (-z) {\boldsymbol e}_z ,  
\, \, e_{z}=(0, 0, 1) .
\end{equation}
The total magnetic charge is zero, while the electric current exists for the Dirac monopole:
\begin{equation}
{\boldsymbol j}^D=\nabla \times {\boldsymbol B}^D=4\pi g \theta(-z)
 [-{\boldsymbol e}_x \delta(x)\delta_(y)'+{\boldsymbol e}_y\delta(x)'\delta_(y)].
\end{equation}
The string singularity exists as a bar magnet {\cite {Boulware1976}}. 

By introducing the pre-potential $ {\boldsymbol C}^D $, the Dirac singularity is derived in a simple 
and consistent way. 

\subsection{The pre-potential method and the singularity for the 't Hooft-Polyakov monopole }  

The magnetic monopole for non-Abelian gauge group $SO(3)$ was discovered by 't Hooft and Polyakov. 
The asymptotic solution for the gauge potential and the scalar field are
 \begin{equation}
{A_i}^{a}=-\frac{1}{e}{\varepsilon}_{iab} \frac{r^b}{r^2}, 
 \hspace{4mm}{\phi}^a=F \frac{r^a}{r},  
 \, \, \, F: \rm{constant} \hspace{4mm} (r\rightarrow\infty) \label{HPsolution}
\end{equation}
where $i, a$ denote space and gauge group indices. 
The electromagnetic field and the covariant derivative for $SO(3)$ 
gauge theory are:  
\begin{equation}
F_{\mu\nu}^{a}=\partial_{\mu}A_{\nu}^{a}-\partial_{\nu}A_{\mu}^{a}
+e\varepsilon^{abc}A_{\mu}^{b}A_{\nu}^{c}, 
\hspace{4mm}
D_{\mu}\phi^{a}=\partial_{\mu}\phi^{a}+e\varepsilon^{abc}A_{\mu}^{b}\phi^{c}. 
\end{equation} 

In the following, we consider the $SU(2)$ gauge group, the covering group of $SO(3)$, 
because of the homotopy group relation: 
\begin{equation}
\pi_2(SU(2)/U(1))=\pi_1(U(1))=Z, 
\end{equation}
where $Z$ stands for the additive group of integer. 
 
We study the magnetic  monopole using our pre-potential method, 
which is very effective for this problem.  
The pre-potential for the 't Hooft-Polyakov monopole is defined to the asymptotic solution as  
\begin{equation}
C_i^{a,HP}=-g\ln r \, \delta_i^a \, ,
\end{equation}
or using the Pauli matrices ($\tau_{x}, \tau_{y}, \tau_{z}$) as 
\begin{equation}
\boldsymbol{C}^{HP}:=-g\ln r\, \boldsymbol{\tau}=-g\ln{r}(\tau_x, \tau_y, \tau_z) . 
\label{Paulimatrices}
\end{equation}
The gauge potential is derived by the curl operation to the pre-potential:
\begin{equation}
\boldsymbol{A}^{HP}=\nabla \times \boldsymbol{C}^{HP}
=-\frac{g}{r^2}(y\tau_z-z\tau_y, z\tau_x-x\tau_z, x\tau_y-y\tau_x) ,
\end{equation}
which is coincident with the HP solution of eq.(\ref{HPsolution}).  
The electromagnetic field strength and the magnetic vector field are evaluated by the pre-potential as  
\begin{align}
B_i^{HP}&=\frac{1}{2}\epsilon_{ijk}F^{HP}_{jk}
=\frac{1}{2}\epsilon_{ijk}(\partial_j A_k^{HP} - \partial_j A_k^{HP}+\frac{e}{2i}[A_j^{HP},A_k^{HP}]) \\
&=(\nabla \times (\nabla \times \boldsymbol{C}^{HP}))_i
+\frac{1}{2}\epsilon_{ijk}
\frac{e}{2i} [(\nabla\times\boldsymbol{C}^{HP})_j,(\nabla\times\boldsymbol{C}^{HP})_k]. 
\end{align}
The first term is calculated using the relation 
$\nabla \times (\nabla \times \boldsymbol{C}^{HP})
=\nabla (\nabla\cdot\boldsymbol{C}^{HP})-\Delta\boldsymbol{C}^{HP}$ as 
\begin{equation}
\frac{1}{2}\epsilon_{ijk}(\partial_jA_k-\partial_kA_j)
=(\nabla (\nabla\cdot\boldsymbol{C}^{HP})-\Delta\boldsymbol{C}^{HP})_i
=2g\frac{r^i\tau^ar^a}{r^3}+g(\Delta\ln r-\frac{1}{r~2})\tau^i . 
\end{equation}
The second term is calculated using the relation $[\tau_{a}, \tau_{b}]=2i \epsilon_{abc}\tau_{c}$  as   
\begin{equation}
\frac{1}{2}\epsilon_{ijk}(\frac{e}{2i}[A_j^{HP},A_k^{HP}]) 
=\frac{1}{2}\epsilon_{ijk}
\frac{e}{2i} [(\nabla\times\boldsymbol{C}^{HP})_j,(\nabla\times\boldsymbol{C}^{HP})_k]. 
=-g\frac{r^i\tau^ar^a}{r^3} .
\end{equation}
Summarizing them, we obtain
\begin{equation}
B^{ia,HP}=g\frac{r^ir^a}{r^4}+g(\Delta \ln r - \frac{1}{r^2})\delta^{ia}. 
\label{HP2.19}
\end{equation}
where the charge quantization relation is applied: 
\begin{equation} eg=-1 . \end{equation}


The gauge independent magnetic field is derived by 
\begin{equation}
B^{i,HP}:=B^{ia,HP}\frac{\phi^a}{|\phi|}
=g\frac{r^i}{r^3}+g(\Delta\ln r -\frac{1}{r^2})\frac{r^i}{r} 
\label{HPmagneticfield}
\end{equation}
The magnetic charge for the first term is the magnetic monopole term: 
\begin{equation}
\rho^{HP}(\rm pole)={\boldsymbol{div}}(g\frac{r^i}{r^3})
=4\pi g \delta^{(3)}(\boldsymbol{r}), 
\end{equation}
and the magnetic charge for the second singular therm is given by 
\begin{align}
\rho^{HP} (\rm singular)&=\partial_i (\Delta \ln r -\frac{1}{r^2})\frac{r^i}{r}
=g\partial_r (\Delta \ln r-\frac{1}{r^2})\notag\\
&=g\Delta \frac{1}{r}=-4\pi g \delta^{(3)}(\boldsymbol{r}) .
\end{align}
The first term and the  second term of the magnetic charge 
have opposite sign and the total magnetic charge becomes zero.    

The magnetic field is therefore re-expressed as the Dirac monopole case as 
\begin{equation}
\boldsymbol{B}^{HP}
=g\frac{1}{r^2}{\boldsymbol e}_{r}
+4\pi g\delta(\bar{x})\delta(\bar{y} \theta(-\bar{z}) )
{\boldsymbol e}_{r}, 
\end{equation}
where the local Cartesian coordinate is used: 
\[d\bar{x}=rd\theta, d\bar{y}=r\sin \theta d\phi, d\bar{z}=dr.\]\\
The first term is the magnetic pole term and the second 
term is 'the hedgehog type singularity'. 

The naive calculation of $(\Delta \ln r -1/r^2)$ leads to zero, but careful 
calculation shows the existence of the singularity. 

\section{Regularization to the pre-potential method}
\setcounter{equation}{0}

In order to confirm the existence of singularities in the Dirac and the 't Hooft-Polyakov monopoles, 
we apply the regularization. 
This is because the singularities appear as the hyper-function 
to which we should pay sufficient attention in analytical calculations. 
The regularization is used to make the hyper-function to be the regular function.   
The small regularization parameter $\epsilon$ is introduced  for 
the radial coordinate in the pre-potential 
\[r\rightarrow r_{\epsilon}=\sqrt{r^2+\epsilon^2}. \] 
and set to zero at the end of calculation. 

\subsection{ Regularization for the Dirac monopole }

The regularized pre-potential for the Dirac monopole is 
\begin{equation}
\boldsymbol{C}_{\epsilon}^D=-g(0,0,\ln (r_{\epsilon}+z)), 
\end{equation}
and the gauge potential is
\begin{equation}
\boldsymbol{A}_{\epsilon}^D=\nabla \times \boldsymbol{C}_{\epsilon}^D
=g\frac{1}{(r_{\epsilon}+z)r_{\epsilon}}(-y,x,0). \label{Dgaugepotential}
\end{equation}
The magnetic field is given by 
\begin{equation}
\boldsymbol{B}_{\epsilon}^D=\nabla \times \boldsymbol{A}_{\epsilon}^D
=\nabla\times(\nabla\times \boldsymbol{C}_{\epsilon}^D)
=\nabla(\nabla\cdot \boldsymbol{C}_{\epsilon}^D)
-\Delta \boldsymbol{C}_{\epsilon}^D
\end{equation}
The first term is calculated to be the pole term as
\[ \boldsymbol{B}_{\epsilon}^D({\rm pole})
=\nabla(\nabla\cdot \boldsymbol{C}_{\epsilon}^D) 
=g{\boldsymbol r}/{r_{\epsilon}^3} .
\]
The magnetic charges for the first pole term and the second singular term become respectively,  
\begin{align}
\rho_{\epsilon}^D(\rm{pole})&=\nabla \cdot 
(g\frac{\boldsymbol r}{{r_{\epsilon}^3}})
\rightarrow 4\pi g \delta^3(\boldsymbol r), \, \\ 
\rho_{\epsilon}^D(\rm{singular})&=\nabla(-\Delta \boldsymbol{C}_{\epsilon}^D)
=g\Delta\frac{1}{r_{\epsilon}}
\rightarrow -4\pi g \delta^3(\boldsymbol r), \, (\epsilon\rightarrow 0).
\end{align}
This indicates that the magnetic field for the second term is 
\begin{equation}
\boldsymbol{B}_{\epsilon}^D (\rm{singular})
=4\pi g \delta(x)\delta(y)\theta(-z)
{\boldsymbol e}_{z}. 
\end{equation}
Therefore the regularization method has established our pre-potential for the Dirac monopole 
of the non-regularization case in section 2.1.  
The result is the same as Boulware et.al., in which their regularization 
starts from the gauge potential as in the same eq.(\ref{Dgaugepotential}) \cite{Boulware1976}. 

\subsection{Regularization for the 't Hooft-Polyakov monopole}

The regularized pre-potential for HP monopole is given by replacing ($r\rightarrow r_{\epsilon}$)
 in eq.(\ref{Paulimatrices}) as
\begin{equation}
\boldsymbol{C}_{\epsilon}^{HP}=-g \ln r_{\epsilon} (\tau_x,\tau_y,\tau_z), 
\end{equation}
and the gauge potential is
\begin{equation}
\boldsymbol{A}_{\epsilon}^{HP}
=\nabla \times \boldsymbol{C}_{\epsilon}^{HP}
=-\frac{g}{ r_{\epsilon}^2}
(y\tau_z-z\tau_y,z\tau_x-x\tau_z,x\tau_y-y\tau_x).
\end{equation}
The regularized magnetic vector field is calculated along the non-regularized case in subsection 2.2 as, 
\begin{eqnarray}
{B}_{i, \epsilon}^{HP}
&=&\frac{1}{2}\epsilon_{ijk}
F_{\epsilon,  jk}^{HP} =\frac{1}{2}\epsilon_{ijk}(\partial_{j}A_{\epsilon, k}^{HP} -\partial_{j}A_{\epsilon, k}^{HP}
+\frac{e}{2i}[A_{\epsilon, j}^{HP},A_{\epsilon, k}^{HP}] ) \nonumber \\
&=& (\nabla \times (\nabla \times \boldsymbol{C_{\epsilon}}^{HP}))_i
+\frac{1}{2}\epsilon_{ijk}
\frac{e}{2i}  [(\nabla\times\boldsymbol{C_{\epsilon}}^{HP})_j,(\nabla\times\boldsymbol{C_{\epsilon}}^{HP})_k]. 
\end{eqnarray}
Following the corresponding calculation to the non-regularization form eq.(\ref{HP2.19}), and we obtain 
\begin{equation}
B_{\epsilon}^{ia,HP}=g\frac{r^ir^a}{r_{\epsilon}^4}+g(\Delta \ln r_{\epsilon} - \frac{1}{r_{\epsilon}^2})\delta^{ia}.
\label{polesingular}
\end{equation}

The regularized colored magnetic charge density is defined as
\begin{align}
\rho_{mag, \epsilon}^{a, HP}=D_{i}B_{\epsilon}^{ia,HP}=\partial_{i}B_{\epsilon}^{ia,HP}
+e\epsilon^{abc} A_{\epsilon}^{ib, HP}B_{\epsilon}^{ic,HP} .
\end{align}
For the first pole term of the magnetic field in eq.(3.10), the magnetic charge density becomes 
\begin{equation}
\rho_{mag, \epsilon}^{a, HP}({\rm pole})=4g\frac{\epsilon^{4}r^{a}}{r_{\epsilon}^{6}} .
\end{equation}  
Using the relation 
\begin{equation}
(\Delta\ln r_{\epsilon}-\frac{1}{r_{\epsilon}^{2}}) =2\frac{\epsilon^2}{r_{\epsilon}^4} , 
\end{equation}
the second singular term of the magnetic field in eq.(3.10) gives 
\begin{equation}
\rho_{mag, \epsilon}^{a, HP}({\rm singular})
= g\partial_{i}(\Delta\ln r_{\epsilon}-\frac{1}{r_{\epsilon}^{2}}) 
+2g(\Delta\ln r_{\epsilon}-\frac{1}{r_{\epsilon}^{2}}) 
=-4g\frac{\epsilon^{4}r^{a}}{r_{\epsilon}^{6}} , 
\end{equation}
The sum of them gives the zero total magnetic charge density.
This confirm the assertion that the total magnetic charge is zero with the hedgehog type singularity 
using our regularization method. 

We note that in the regularized scalar field solution $\phi_{\epsilon}^a $ cannot be obtained 
because the scalar field equation does not satisfied:   
\begin{equation}
D_i\phi_{\epsilon}^a=\partial \phi_{\epsilon}^a+e\epsilon^{abc}A_{\epsilon}^{ib}\phi_{\epsilon}^c
\neq 0 , 
\end {equation}
for a given regularized gauge potential $A_{\epsilon}^{ib}$ in eq.(3.8).
Therefore the regularized magnetic field $B_{\epsilon}^{i,HP}=B_{\epsilon}^{ia,HP}{\phi^a}/|{\phi}|$
cannot be gauge invariant and 
we derive the colored magnetic charge density 
without using the gauge invariant regularized form.  
.  

\section{SU(2) gauge transformation between the Dirac and the  't Hooft-Polyakov monopoles}
\setcounter{equation}{0}

In this section, we consider the SU(2) gauge transformation between the Dirac monopole 
and the 't Hooft-Polyakov monopole in order to study their singularities. 
For this purpose, the Dirac monopole theory is embedded in SU(2) gauge theory in the third direction.  
The gauge potential and the scalar field are given as 
\begin{align}
{\boldsymbol{A}^D}=\frac{g (-y, x, 0)}{r(r+z)}\tau_z \,\, , \, \, \,  
\phi^D&=F\tau_z \,\,\,\, \,  (F=\rm{constant}) .
\end{align}

By the SU(2) gauge transformation from ${z}$ to $\vec{r}$ direction 
(polar angle $\theta$, azimuthal angle $\phi$) 
\begin{equation}
S=\exp{(-i\phi \tau_3/2)\exp(i\theta \tau_2 /2) \exp(i\phi \tau_3 /2)}.,  
\end{equation}  
the gauge potential and the scalar field are transferred as 
\begin{align}
S\boldsymbol{A}^{D}S^{-1}+\frac{2i}{e}S\nabla S^{-1} 
&=-\frac{1}{er^2}(y\tau_z-z\tau_y, z\tau_x-x\tau_z, x\tau_y-y\tau_x) ] \\ 
S\phi^{D}S^{-1}&=F\frac{r^a\tau^a}{r} , 
\end{align}
which coincide with the 't Hooft-Polyakov monopole solution $\boldsymbol{A}^{HP}$ and $\phi^{HP}$ 
in eq.(\ref{HPsolution}) respectively  \cite{Ryder}. 

The existence or non-existence of singularity in the monopole theories is very important 
during the gauge transformation. 

In this gauge transformation, the polar angle regularization method was introduced 
by Boulware et. al \cite{Boulware1976}: 
\begin{equation}
\theta_{\epsilon}:
=\theta \, \frac{1+\cos \theta}{1+\cos \theta +\epsilon^2}
\end{equation}
and obtained the result that the Dirac string singularity disappeared. 
However, the singularity position of $\theta=\pi$ is obtained both of $(-z)$ and $(-r)$ directions 
according to take the observer's standard coordinate system.    

As another investigation, the transformation from $z$ to $\vec{r}$ direction becomes singular for 
\begin{equation}1+\boldsymbol{e}_z\cdot \boldsymbol{e}_r=
1+\cos \theta=0 , \end{equation}  
and the singularity thought to exist in $(-z)$ direction \cite{Weinberg}. 
In this criterion, singularities are appeared in $(-z)$ direction as well as in $(-\bar{z})$ direction 
in the local Cartesian coordinate, according to our analysis with the pre-potential method.

\section{Conclusion and remarks}
By using the pre-potential method, we can analyze the magnetic poles and singularities in parallel manner 
for the Dirac and the  't Hooft-Polyakov monopoles. 
The regularization method supports our result. 
In the gauge transformation from Dirac to 't Hooft-Polyakov monopoles, 
the angular singularity ($\theta = \pi $) i.e. singularity along ($-z$  )
was considered already  
but the radial singularity along ($-\bar{z}$) in the local Cartesian coordinate also exists. 
Therefore the gauge transformation changes the singularity position but
conserves the magnetic charge (which is zero) and their energy (which is infinity).   

As a result, 
the monopoles do not exist for both of the Dirac and the 't Hooft-Polyakov magnetic monopole theories but 
magnetic singularities and the result is consistent with the no magnetic monopole observation 
until now (see for example, Magnetic Monopole Searches in Particle Data Group \cite{particledata}). \\

Some remarks are as follows. \\
\, \, \, (1) Bogomol'nyi, Prasad and Sommerfeld found the analytic solution to approach the 't Hooft-Polyakov asymptotic 
solution (2.9), which is called as BPS monopoles \cite{Bogomol'nyi} \cite{PrasadSommerfeld}:
\begin{equation}
A_i^a=\epsilon_{aij}\frac{r^a}{er^2}(1-\frac{\rho}{\sinh \rho}) , \, \, \phi^a=\frac{r^a}{er^2} (\coth\rho-\frac{1}{\rho}) , 
\end{equation} 
with $\rho:=er|\phi|$. 
The magnetic charge is obtained by the surface integral of the magnetic field, which is determined by 
the asymptotic behavior of the gauge potential and the scalar field.  
And our analysis in sec. 2.2 and sec. 3.2 is satisfied and the conclusion is unchanged. 
\\

(2) It is worthwhile to note on the Lipkin, Weisberger and Peshkin difficulty \cite{Lipkin1969}. 
They state that the Jacobi identity  does not hold for the kinetic momentum operator $\pi_i=p_i-e A_i$ 
 in the magnetic monopole case as
\begin{equation}
[[\pi_1,\pi_2],\pi_3]+[[\pi_2,\pi_3],\pi_1]+[[\pi_3,\pi_1],\pi_2]\neq 0 , 
\end{equation}
and state that the contradiction could be removed to exclude the origin. 
This difficulty is removed by introducing the magnetic singularities 
for both the Dirac and the 't Hooft-Polyakov monopoles.   
\\

(3) 
The gauge independent electromagnetic tensor field is proposed by 't Hooft \cite{'t Hooft1974}  
\begin{equation}
\textsf{F}_{ij}:=
F_{ij}^a\frac{\phi^a}{|\phi|}-\frac{1}{e|\phi|^3}\epsilon_{abc}\phi^a(D_i\phi^b)(D_j\phi^c) . 
\label{Htensor}
\end{equation}
The first term contributes to  the asymptotic solutions both of the Dirac and the 't Hooft-Polyakov monopoles,
and is estimated by using the pre-potential method in sec.2 and sec.3. 
This gauge independent tensor is rewritten 
using the identity $\epsilon_{abc}\epsilon_{ade}=\delta_{bd}\delta_{ce}-\delta_{be}\delta_{cd}$ in the form 
\begin{equation}
\textsf{F}_{ij}=(\partial_{i}A_j - \partial_{j}A_i) 
-\frac{1}{e|\phi|^3}\varepsilon_{abc}\phi^a(\partial_i\phi^b)(\partial_j\phi^c) , \label{AFGtensor}
\end{equation}
with  the definition $A_i:=A_i^a \phi^a/|\phi|$. 
The first term contributes to the Dirac monopole. 
The second term contributes to the 't Hooft-Polyakovc topological monopole 
and is explained as the Brouwer degree of the mapping from coordinate space to isospin space 
 by Arafune, Freund and Goebel \cite{Arafune1975}. 
However the derivation from eq.(\ref{Htensor}) to eq.(\ref{AFGtensor})  takes account of 
only the magnetic pole term $gr^i/r^3$ in eq.(\ref{HPmagneticfield}).  
A careful calculation by the pre-potential method shows the existence of singularities 
and the non-existence of magnetic monopoles. \\

Our pre-potential analysis is consistent with the standard Maxwell's electromagnetic theory.

\section*{Acknowledgements}

The author would like to thank Prof. Yong-min Cho for his invitation to the monopole research.  The author also gives thanks to Prof. Kazuyasu Shigemoto for his helpful suggestions and comments on this work. 


\end{document}